\documentclass[adp,fleqn,referee]{w-art}
\usepackage{times}
\usepackage{amsmath}
\usepackage{w-thm}
\usepackage{bm}

\chardef\bslash=`\\ 

\newcommand{\eval}[2][\right]{\relax
  \ifx#1\right\relax \left.\fi#2#1\rvert}

\def\a{\alpha}
\def\be{\beta}

\def\w{\omega}

\def\Vac{\Omega}

\def\e{\epsilon}

\def\s{\sigma}

\def\de{\delta}

\def\th{\theta}

\def\vp{\varphi}

\def\d{\partial}

\def\v#1{{\mathbf{#1}}}
\def\h#1{#1^*}

\def\b#1{\overline{#1}}

\def\*{\cdot }

\def\ra{\rightarrow}

\def\H{\mathcal{H}}

\def\Fou{\mathcal{F}}

\def\ess_sp{\s_{ess}}
\def\F1{\Fou^{(1)}}

\def\R{\mathbf{R}}
\def\CC{\mathbf{C}}

\begin{document}
\Volume{XX}
\Issue{1}
\Month{1}
\Year{2003}
\pagespan{3}{}
\Receiveddate{25 November 2004}
 \Accepteddate{15 March 2005 by U. Eckern}




\title[Spontaneous emission of light]{Spontaneous emission of light from atoms:  the model}


\author[P. Marecki]{P. Marecki\footnote{Corresponding author E-mail: {\sf piotrm@wsi.edu.pl}}\inst{1}}
\address[\inst{1}]{Wyzsza Szkola Informatyki i Zarzadzania, Bielsko-Biala, Poland}

\author[N. Szpak]{N. Szpak\inst{2}}
\address[\inst{2}]{Institute for Theoretical Physics,
J.W. Goethe University, Frankfurt/Main, Germany}

\begin{abstract}
We investigate (non-relativistic) atomic systems interacting with quantum
 electromagnetic field (QEF). The resulting model describes spontaneous
emission of light from a two-level atom surrounded by various
initial states of the QEF. We assume that the quantum field
interacts with the atom via the standard, minimal-coupling
Hamiltonian, with the $A^2$ term neglected. We also assume that
there will appear at most single excitations (photons). By
conducting the analysis on a general level we allow for an
arbitrary initial state of the QEF (which can be for instance:
the vacuum, the ground state in a cavity, or the squeezed state).
We derive a Volterra-type equation which governs the time
evolution of the amplitude of the excited state. The two-point
function of the initial state of the QEF, integrated with a
combination of atomic wavefunctions, forms the kernel of this
equation.
\end{abstract}

\maketitle

\renewcommand{\rightmark}
{\textsc{}}

\section{Introduction}
The phenomenon of emission of photons during the transition
between atomic energy-levels is one of the most important, and
well understood aspect of contemporary theoretical physics. The
explanations of this process which have already been presented in
the early days of quantum theory \cite{WW}, vary in their level
of sophistication. The desirable formulation, in the spirit of
quantum theory, describes the process as a unitary evolution of
the full system which is composed of the atom and the quantum
electromagnetic field. It is disappointing that modern literature
in most cases reports models which either do not fit to this
basic requirement of quantum theory, or at some point involve
unjustified operations with (unnecessarily) singular
expressions\footnote{See, for instance, \cite{SZ} chapter 6.3.}.
One has an unpleasant feeling that the outcome (the exponential
decay of the excited-state's amplitude) is simply assumed at some
point of the investigations. While we do not claim that no clean
model exists, we believe a simple field-theoretical model which
is additionally more general with respect to the quantum
electromagnetic field (QEF), is worth presenting.

It is generally recognized that the atomic emission of radiation
is influenced by both: the atomic structure (in particular: the
energies and wavefunctions of the levels under consideration) and
the initial state of the QEF (which can be a strong, coherent
state or a vacuum, for instance).

With respect to the latter we note that there exists a great
variety of states of QEF. On the level of local quantum field
theory \cite{haag}, it is known that even disjoint representations
of the algebra of fields (the Canonical Commutation Relations
(CCR)-algebra) can be constructed\footnote{It is possible, for
instance, to prescribe two states of the QEF which cannot be
expressed as density-operator states in a common Fock space.}.
Each such representation carries its own notion of excitations
(``photons''). A general method of Gelfand, Naimark and Segal
(GNS) allows to construct a representation once a single
``reference state" $\Vac$ is specified. If we deal with fields
propagating on the Minkowski spacetime, the Fock representation,
based on the usual vacuum, is physically distinguished. However,
if we were interested in some generic spacetimes (or
time-dependent environments) there would be no privileged
``reference state''. On the mathematical level, one is content
with a picture of many permissible representations, each based on
a different ``reference state'' $\Vac$, although such a state of
affairs is certainly conceptually worrying  (there is no unique
answer to the question what a photon is).

On the other hand, experimental advances in the area of quantum
optics have produced a variety of states of light which exhibit
non-classical features. The so-called squeezed states or the
ground states of the QEF in small cavities provide environments
which are much different form the usual Minkowski-spacetime
vacuum. In particular, the experiments with atoms surrounded by
squeezed states,  \cite{kimble},  revealed a modification of
their spectroscopic characteristics.

One of the simplest ways to test the properties of various states
of QEF is to let the quantum field interact with a quantum system
of relatively simple structure. As an example, we can consider a
``two-level atom'', assume a simple minimal-coupling interaction
with the QEF, and investigate the time-evolution of various
initial states of the full system. Such an approach is also
experimentally viable, as this evolution is nowadays fully
accessible to measurements (see e.g. \cite{holography}). In order
to illustrate the potential of such an approach we remark on a
possibility to measure the Unruh effect with the help of an
electron, namely, a simple perturbative calculation shows that
the state of a two-level quantum system coupled to the QEF
initially in the KMS (thermal) state thermalizes to the
temperature of the KMS state. As originally suggested by  Bell
and Leinaas \cite{bell}, a spin of a linearly-accelerated
electron might serve as a detector of the Unruh
temperature\footnote{The usual vacuum state, when observed by an
accelerated observer, appears to be a thermal state, with an
acceleration-dependent temperature.}. It is one of the aims of
this paper to pursue further this type of models in which test
atomic systems are employed as detectors of properties of various
states of the QEF.

The paper contains a field-theoretical model of an interaction of a
two-level atom with the QEF. We assume the atom to be initially in an
excited state, and let it interact with the QEF; the dynamics of such a
system is usually referred to as the \emph{spontaneous emission of
radiation}. It has been investigated in the past by many authors with
various emphasis on mathematical rigor, on the one hand, and on physical
concreteness on the other. We shall briefly summarize these attempts
here, in order to put our model into a context.

 The simplest way to estimate the dynamics of the full system is to use
the time-dependent perturbation theory w.r.t. the interaction of
the QEF and the atom (Eq. \eqref{equ:interaction}). By doing so,
one can approximate the initial dynamics of the emission of a
photon, however, the long-term behavior of this process cannot be
estimated in this way. Another method, due to Weisskopf and
Wigner \cite{WW}\footnote{See, for instance, the textbook of
Scully and Zubairy \cite{SZ}.}, establishes an ordinary
integro-differential equation for the time-dependence of the
amplitude of the excited state. The model we shall present here
generalizes the Weisskopf-Wigner model with respect to the allowed
initial states of the quantum radiation field.

With an alternative approach,  Bach and collaborators \cite{bach}
have investigated systems of non-relativistic particles coupled
to the QEF from the functional-analytic point of view. Without
any assumption on the number of atomic energy levels, they worked
with the full minimal coupling Hamiltonian (together with the
quadratic term, and an ultraviolet cutoff), and have been able to
prove the existence of a ground state of the coupled system.
Moreover, an estimate has been derived for the life-times of the
states of the atom which were stationary when the interaction was
absent.

A common feature of all these developments is, that the choice of the
initial state of the QEF is made at a very early stage of investigations.
Consequently, given the final equation for the excited state's amplitude,
it is difficult to discern which features of this equation are related to
the initial state of the QEF, and which are influenced by the atomic
wavefunctions\footnote{In an interesting paper Bondarev and Lambin
\cite{bon}, investigate the spontaneous emission of light from atoms
placed inside of small cavities (carbon nanotubes). Their explicit
numerical results complement our analysis and exemplify the possible
complexity of the emission process.}.

In this paper we attempt to generalize the Weisskopf-Wigner
model; in particular we proceed in a general way, so that the
initial state of the QEF can be left unspecified to the very end.
By doing so, we prove that the spontaneous emission process is
governed by a Volterra integro-differential equation (Eq.
\eqref{main_eq}), the kernel of which is composed of the
two-point function, $\bigl(\Vac,A_i(x) A_j(y) \Vac\bigr)$, of the
initial state of the radiation field, smeared with test functions
which are derived from the wavefunctions of the ground- and
excited-state of the atom. As the two-point functions are the
foundation of the GNS construction, and on the other hand are
known for physically interesting states, this opens up a
possibility to investigate various initial states of radiation on
an equal basis.

\section{Formulation of the model}
\subsection{General setting}

A natural Hilbert space for a system
of an atom (non-relativistic) coupled to the quantum radiation
field is
\begin{equation}
    \H=L^2(\R^3)\otimes \Fou,
\end{equation}
where $\Fou$ denotes the physical (transversal) Fock space of the
radiation field. This space will be later restricted to a subspace,
consisting of a two-dimensional Hilbert space of atomic states,
tensor-multiplied with the vacuum- and one-excitation-subspace of $\Fou$.
The single excitations (over the reference state $\Vac$) of the quantum
radiation field, denoted by the symbol $|f\rangle$, will be described by
\begin{equation}
|f\rangle=A(f)\Vac=\int d^3x\ f_i(\v x) A^i(0,\v x)\Vac,
\end{equation}
where $A_i(0,\v x)$ is the (unsmeared) vector-potential operator at
$t=0$. The complex-valued test function $f_i(\v x)$ will carry the
information about the "wave packet" of the excitation\footnote{If $\Vac$
is the vacuum on the Minkowski spacetime, the excitations are photons. In
this case $A(f)\Vac=\int \frac{d^3p}{\sqrt{2p}}\ f^\a(\v p)a_\a^*(\v p)\
\Vac$, where $\h a_\a(\v p)$ is the creation operator for $\a$-th
polarization and $f^\a(\v p)=\int d^3x\ f^i(\v x) \e_i^\a(\v p) e^{-i\v
p\v x}$. The symbols $\e_i^\a(\v p)$ denote the transversal polarization
vectors.}.

The interaction between the atom and the QEF will be generated by
\begin{equation}\label{equ:interaction}
  V(t,\v x)=-\sqrt{\a}\  A^i(t,\v x)\*\v p_i,
\end{equation}
where\footnote{In the paper we use dimensionless units (see
appendix \ref{appa}). By the letter $p$ we denote the frequency:
$p=|\v p|$.} $\v p_i=-i \d_i$, $\a$ is the fine-structure
constant, and $A^i(t,\v x)$ denotes the vector-potential
operator, in the radiation gauge, with the free time-evolution
already implemented\footnote{Which means that the free
time-evolution of $A^i(t,\v x)$ is governed by the Maxwell
equations.}. This is a standard interaction Hamiltonian with the
$A^2$ term neglected.

\subsection{Dynamics of the restricted system}
In this section we will consider the evolution of the above
system in the situation where the radiation field is initially in
the state $\Vac$, about which we only assume that the
\emph{expectation value of an odd number of field operators in
this state vanishes}\footnote{The vacuum, the ground state, the
squeezed state and a general class of quasi-free states fulfill
this requirement.}, and the atom is initially excited.

The Hilbert space of the atomic motion will be
restricted to a two-dimensional subspace (two energy levels), and the Fock space will be restricted to the zero- and one-excitation-subspace\footnote{Every vector
$v$ of such a space can be written as: $v= c\Vac+  A(f)\Vac$ with a
complex number $c$ and complex, vector-valued functions $f^i(\v x)$.}.

We start by restricting the interaction \eqref{equ:interaction} to the
subspace of only two energy-levels. To do so, we substitute the
interaction operator $V$ by $V_2$:
\begin{align*}
  V_2 &=\sum_{ij=0,1} (\psi_i,V\ \psi_j)_{L^2}\ |\psi_i\rangle\langle\psi_j|,\\
  (\psi_i,V\ \psi_j)_{L^2} &=\int d^3x\ \b{\psi_i(\v x)} V \psi_j(\v x),
\end{align*}
where we have employed the Dirac's notation. We obtain\footnote{The terms
$|\psi_1\rangle\langle\psi_1|$ and $|\psi_0\rangle\langle\psi_0|$ drop
out, if the wavefunctions are real (which we hereby assume),  because
$\int \b{\psi_0}(\d_i\psi_0) A^i=0$, as a consequence of the chosen gauge
condition $\d_iA^i=0$.}:
\begin{equation}
V_2=i\sqrt{\a}\bigl[|\psi_1\rangle\langle\psi_0|\otimes A_t(\chi) + |\psi_0\rangle\langle\psi_1|\otimes
  A_t(\vp)\bigr],
\end{equation}
where
\begin{align*}
\chi_i(\v x)&= \b \psi_1(\v x)\  \d_i \psi_0(\v x),\\
 \vp_i(\v x) &= \b
\psi_0(\v x)\ \d_i \psi_1(\v x)\\
A_t(\chi)&=\int d^3x\ A_i(t,\v x)\chi^i(\v x).
\end{align*}
Because of $A_t(\chi)=-A_t(\vp)$, the interaction can be written
as
\begin{equation}
  V_2=\sqrt{\a}\ \sigma_2\otimes A_t(\chi),
\end{equation}
which is a selfadjoint operator on $\CC^2\otimes\Fou$. At this
point we remark that the process of ``spontaneous emission of
radiation'' will be generated by a selfadjoint Hamiltonian, and
thus will be governed by a unitary evolution. In this way the
basic requirement of the quantum theory will be fulfilled.

We shall now additionally restrict the investigations to the
subspace of at most one excitation of the QEF\footnote{This
restriction allows us to derive a single equation for the
evolution of the excited-state's amplitude (Eq. \eqref{main_eq}).
We expect such a restriction to be justified in some cases, e.g.
where one-photon transitions are known (from experiments) to be
dominant.}. Later, we will employ the interaction picture and
derive a system of evolution equations. At the time $t$, the
state of the full system, in the restricted space, can always be
parameterized by two complex, vector-valued functions $f_t(\v
x),g_t(\v x)$ (excitation wavepackets), and two complex numbers
$c(t),d(t)$:
\begin{equation}
  S=c(t)\ \psi_1\otimes \Vac+d(t)\ \psi_0\otimes \Vac+\psi_0\otimes
A_0(f_t)\Vac+\psi_1\otimes A_0(g_t)\Vac.
\end{equation}
We note that we speak of excitations because the state $\Vac$ is
still unspecified. We use the following initial conditions
\begin{align*}
  c(0) &=1, \\
  f_0(\v x),g_0(\v x),d(0) &=0.
\end{align*}
For such conditions the functions $g_t(\v x)$ and $d(t)$ never
acquire non-zero values, and will be omitted from now on. In the
interaction picture, we reparameterize the state of the full
system as follows:
\begin{equation}
S_I(t)=c(t)\ \psi_1\otimes \Vac+  \psi_0\otimes A(f_t)\Vac.
\end{equation}
The time evolution of $c(t)$ and $f_t$ needs to be determined; it
will be generated by the operator $V^I_2$, which is the operator
$V_2$ in the interaction picture:
\begin{equation}
  V^I_2= i\sqrt \a\left(e^{i\w t}|\psi_1\rangle\langle\psi_0| -
  e^{-i\w t}|\psi_0\rangle\langle\psi_1|\right)\otimes
  A_t(\chi),
\end{equation}
where $\w=E_1-E_0$ is the energy difference of both states of the atom.
We shall introduce a decomposition of $f_t$ in terms of an orthonormal
basis $\{f_m\}$ of the one-excitation space
\begin{equation}
  A(f_t)\Vac=\sum_{m=1}^\infty c_m(t)A(f_m)\Vac,
\end{equation}
where $f_m$ are complex functions, orthonormal w.r.t. the scalar
product\footnote{Here, the two-point function of the initial state of the
QEF, $\w_{ij}(t,\v x,s,\v y)=\bigl(\Vac,A_i(t,\v x) A_j(s,\v y)
\Vac\bigr)$, enters the investigations. The one-excitation space is
separable for reasonable initial states $\Vac$.}
\begin{equation}
(f,g)=\int {d^3x}\ \w_{ij}(0,\v x,0,\v y) f^i(\v x) g^j(\v y).
\end{equation}

The evolution equation,
\begin{equation}
i\frac{dS_I}{dt}(t) = V^I_2(t)\ S_I(t),
\end{equation}
after taking scalar products with $\psi_1\otimes\Vac$ and
$\psi_0\otimes A(f_n)\Vac$, leads to an infinite system of ordinary differential equations:
\begin{subequations}
\begin{align}
i\dot c(t)&=i\sqrt\a\ e^{i\w t}\*\sum_{m=1}^{\infty} c_m(t)\ (\Vac,A_t(\chi)A(f_m)\Vac),\label{pierwsze}\\
i\dot c_n(t)&=-i\sqrt\a\ e^{-i\w t}\* c(t)\ (\Vac, A(f_n)A_t(\chi)\Vac)\label{drugie}.
\end{align}
\end{subequations}
Integrating the second equation(s), \eqref{drugie}, from $0$ to $t$,
with the initial condition $c_n(0)=0$ (for all $n$), yields
\begin{equation}
  c_n(t)=-\sqrt \a \int_0^tds\ e^{-i\w s} c(s)\ (\Vac, A(f_n)A_s(\chi)\Vac).
\end{equation}
After substituting this into the equation for $c(t)$, \eqref{pierwsze}, we obtain
\begin{equation}
\dot c(t)=-\a\int_0^tds\ c(s)\ e^{i\w (t-s)}\sum_{m=1}^{\infty}  (\Vac,
A(f_m)A_s(\chi)\Vac)\ (\Vac, A_t(\chi)A(f_m)\Vac),
\end{equation}
and finally, as a consequence of the completeness of the $\{f_m\}$ in the
one-excitation subspace of $\Fou$, we get
\begin{equation}\label{main_eq}
\dot c(t)=-\a \int_0^tds\ c(s)\ e^{i\w (t-s)}
(\Vac,A_t(\chi)A_s(\chi) \Vac)=-\a \int_0^tds\ c(s) e^{i\w(t-s)}\
S(t,s).
\end{equation}
with
\begin{equation}
  S(t,s)=\int d^3x\ d^3y\ \w_{ij}(t,\v x,s,\v y)\chi^i(\v
  x)\chi^i(\v y).
\end{equation}

The equation \eqref{main_eq}, valid for arbitrary initial states of the
QEF, is a Volterra integro-differential equation of the second type. In
the important case of a \emph{stationary initial state} $\Vac$, the
smeared two-point function $\w(t,\chi,s,\chi)$ depends only on the
difference\footnote{As a consequence, $S(t,s)=S(t-s,0)$.}, $t-s$, and
therefore in this case we obtain a convolution Volterra
integro-differential equation. Such an equation can be brought into the
form of an integral equation. To show this, we integrate it w.r.t. the
variable $t$ from zero to $T$ and (with an appropriate change of
variables) obtain
\begin{equation}
    c(T)=1-\int_0^T Z(T-s) c(s)\ ds, \qquad
\end{equation}
where
\begin{equation}
    Z(\tau)=\a \int_0^\tau dt\ S(t,0) e^{i \w t}.
\end{equation}

An alternative way to approach the  equation \eqref{main_eq}, in the case
of stationary initial states, is to perform a Laplace transform of it.
For the transformed amplitude, $\hat c(s)$, we obtain a relation
\begin{equation}
    \hat{c}(s) = \frac{1}{s+\alpha \hat{S}(s-i\omega)},
\end{equation}
where $\hat{S}(s)=\int_0^\infty \exp(-st)\ S(t,0)\ dt$. The
inverse Laplace transform gives the original amplitude which we
intend to compute:
\begin{equation}
c(t) = \frac{1}{2\pi i} \int_C \frac{e^{st}}{s+\a\hat{S}(s-i\omega)}\; ds,
\end{equation}
where the contour of integration, $C$, needs to be chosen parallel to the
imaginary axis, to the right of all poles. In our case, the dynamics of
the total system is unitary, and consequently there will be no poles of
the integrand for $Re(s)>0$. The standard procedure is to attempt to
close the contour $C$ from the left ($Re(s)<0$). The dominant long-term
asymptotic behavior of $c(t)$ comes from the singularities of $\hat c(s)$
with the greatest $Re(s)$. In general, the integrand, analytically
continued to $Re(s)<0$, may exhibit branch points (which often lead to
power-law decays) and simple poles (which lead to exponential decays). It
is not possible to be any more specific here, without specifying $\Vac$,
as the analytic properties $\hat{S}(s-i\omega)$ obviously depend
crucially on it.

In any case, even for non-stationary initial states, the kernel
of the equation \eqref{main_eq} is a smooth function of its
arguments (because the test functions $\chi$ are smooth), and
therefore our problem always reduces to solving \emph{a
Volterra-type integro-differential equation of the second kind
with a smooth kernel}. The numerical experience with this
equation, for various initial states, shows that the solutions
exhibit a non-trivial initial dynamics which is followed by a
relatively uncomplicated, monotonic-decay phase\footnote{If the
atom is placed inside of a reflecting cavity, however, $c(t)$
exhibits a completely different dynamics, due to the interaction
with the emitted photon. In an extreme case this can even lead to
oscillations of $|c(t)|^2$ which are called the vacuum Rabi
oscillations \cite{bon,one}.}.

We wish to defer a systematic study  of the solutions of \eqref{main_eq}
to a future publication \cite{pinq2}.

\subsection{Special case: Minkowski vacuum as the initial state $\Vac$}
The simplest case of a great physical importance arises, if one takes the
usual vacuum state $\Vac$ as the initial state of QEF.  Then, the
electromagnetic field operator can be expressed in terms of the creation
and annihilation operators,
\begin{equation}\label{A}
     A_i(t,\v x)=\frac{1}{\sqrt{2\pi}^3} \int \frac{d^3k}{\sqrt{2k^0}}
  \  \e^\a_i(\v k)
  \left\{ a^*_\a (\v k)\ e^{ikx}+a_\a (\v k)\ e^{-ikx}\right\},
\end{equation}
where $\e^\a_i(\v k)$ denote the two ($\a=1,2$) $\v k$-dependent, transversal
polarization vectors, and the vacuum is defined via
\begin{equation}
a_\a (\v k)\Vac=0.
\end{equation}
The respective two-point function is well-known:
\begin{equation}
\w^0_{ij}(t,\v x,s,\v y)=(\Vac,A_i(t,\v x) A_j(s,\v y)\ \Vac)=
\frac{1}{(2\pi)^3}\int\frac{d^3p}{2p}\ e^{-ip(t-s)}\left(\de_{ij}-\frac{\v
p_i\v p_j}{\v p^2}\right) e^{i\v p(\v x-\v y)}.
\end{equation}
Denoting by $\chi_i(\v p)$ the Fourier transform of $\chi_i(\v x)$,
\begin{equation}
  \chi_i(\v p)=\int d^3x\ e^{-i\v p\v x}\ \b \psi_1(\v x) \d_i \psi_0(\v x),
\end{equation}
we get
\begin{equation}\label{vac_2_point}
  S(t-s)=\frac{1}{(2\pi)^3}\int\frac{d^3p}{2p}\ e^{-ip(t-s)}\left(\de^{ij}-\frac{\v p^i\v p^j}{\v p^2}\right)\
  \b{\chi_i(\v p)} \chi_j(\v p)
\end{equation}
The function $S(t-s)$ is evidently a smooth function of its
arguments\footnote{In particular, the limit $s\rightarrow t$ is
finite.} , because $\chi(\v p)$ is a smooth function of rapid
decay\footnote{The smoothness of $\chi(\v p)$ assures, that the
point $p=0$ is an integrable singularity. The decay property for
large $|\v p|$ makes the integral convergent for large momenta,
even when the time derivatives of $S(t-s)$ are taken.}.

\subsection{Another special case: squeezed state as the initial state $\Vac$}
The squeezed states are examples of states of the QEF which can be
produced experimentally \cite{SZ}, and exhibit exotic, inherently
quantum properties\footnote{Such as, for instance, a sub-vacuum
level of electric field fluctuations.}. These states appear as a
coherent superposition of pairs of photons. The photons
constituting a pair, in the case of a degenerate squeezed state,
possess the same wavepacket, $f_\a(\v p)$. (We assume this
function to be  smooth and rapidly decaying.) The squeezed states
are important in the context of this paper, as their presence is
known (experimentally) to influence the atomic dynamics
\cite{kimble}.

A squeezed state, $|S\rangle$, is a vector state created in the
Fock space constructed upon the (distinguished,
Minkowski-spacetime) vacuum with the help of a unitary operator
$S(f,r)$:
\begin{equation}
|S\rangle = S(f,r)\Vac=e^{ r\left[ a(f)a(f)- a^*(f)a^*(f)\right]} \Vac,
\end{equation}
where $r$ is a real number which we interpret as the amplitude of
squeezed light\footnote{As the arguments of $S(f,r)$ will not be
varied in the following discussion, we will briefly write $S$
instead of $S(f,r)$.}. The operator $a(f)$ denotes here the
smeared annihilation operator\footnote{The index $\a$ is summed
over, $\a=1,2$, whenever it appears twice.}:
\begin{equation}
a(f)=\int d^3p\ \b{ f_\a(\v p)} a^\a(\v p).
\end{equation}
The following (commutation) relations hold for an arbitrary test function
$g(\v p)$, and facilitate the calculation of various expectation values:
\begin{align}
S^* a(g) S&=a(g)-(g,f)\ a(f)+(g,f)\left[ a(f)\cosh r - a^*(f)\sinh r \right]\\
S^* a^*(g) S&=a^*(g)-(f,g)\ a(f)+(f,g)\left[ a^*(f)\cosh r - a(f)\sinh r \right],
\end{align}
where $(f,g)=\int d^3p\ \b{f_\a(\v p)} g^\a(\v p)$ stands for the scalar
product of $f^\a$ and $g^\a$. With the help of these  relations we find
\begin{align*}
\left( S\Vac, a^*_\a(\v p)a_\be(\v k)\ S\Vac\right) &=\b{f_\a(\v p)}f_\be(\v k) \sinh^2 r\\
(S\Vac, a_\a(\v p)a_\be(\v k)\ S\Vac)&=-f_\a(\v p)f_\be(\v p) \sinh r \cosh r
\end{align*}
The two-point function of the radiation field, in the squeezed state, can
now be computed (here $x\equiv (x_0,\v x)$)
\begin{multline*}
\w^S_{ij}(x,y)=(\Vac,A_i(x) A_j(y)\ \Vac)=\w^0_{ij}(x,y)+
\int\frac{d^3p\ d^3k }{2\sqrt{pk}}\e^\a_i(\v p)\e^\be_j(\v
k)\*\\\* \left\lbrace - \Bigl[f_\a(\v p) f_\be(\v k)e^{-ipx-iky}
+{\rm c.c.} \Bigr]\sinh r\cosh r+
 \Bigl[\b{f_\a(\v p)} f_\be(\v k)e^{ipx-iky} +{\rm c.c.} \Bigr]\sinh^2 r\right\rbrace .
\end{multline*}
Evidently, it differs from the vacuum-two-point function,
$\w^0_{ij}(x,y)$, by a function which is smooth in both arguments.

If the wavepacket $f_\a(\v p)$ is concentrated around a specific momentum
$\v q$, then the two-point function of the squeezed state simplifies to
\begin{multline*}
\w^S_{ij}(x,y)=\w^0_{ij}(x,y)- \left\lbrace  \Bigl[d_i d_j
e^{-iq(x+y)}+{\rm c.c.} \Bigr]\sinh r\cosh r+
 \Bigl[\b{d_i} d_j e^{iq(x-y)}+{\rm c.c.} \Bigr]\sinh^2 r\right\rbrace,
\end{multline*}
where $\v d$ is a polarization vector (orthogonal to $\v q$).
From the above expression one realizes that the difference
between $\w^S$ and $\w^0$ becomes negative for some periods of
time. This fact corresponds to the sub-vacuum fluctuations of the
vector potential\footnote{The square of $A^i(x)$ is defined via
normal ordering w.r.t. the vacuum $\Vac$.} $A_i$; the expectation
values of the squared electric-field operator, and the
energy-density operator behave similarly. We note that the
squeezed state is not a stationary state, and therefore the
Volterra equation \eqref{main_eq} is will not be a
convolution-type equation in this case.

\section{Example}

\subsection{Hydrogen atom, $2P\ra 1S$ transition}
The wavepackets of the initial and final state of the atom are
\begin{align*}
  \psi_0&=\frac{\a^{3/2}}{\sqrt{\pi}}\ e^{-\a r},&
  \psi_1&=\frac{\a^{5/2}}{4\sqrt{2\pi}}\ r \cos\th\   e^{-\a r/2}.
\end{align*}
We will compute the smearing functions,
\begin{equation}
\chi_i(\v p)=\int d^3x\ e^{-i\v p\v x}\ \b \psi_1(\v x)\v\ \d_i \psi_0(\v
x).
\end{equation}
The differentiation w.r.t. $x^i$ leads to a factor $x^i$
(multiplied with functions of $r$ only) which can be expressed as
a derivative $i\d/\d p_i$. Similarly, the term $z=r \cos(\th)$,
can be expressed as a derivative w.r.t. the $p_z$. In this way we
obtain
\begin{equation}
\chi_i(\v p)=\frac{\a^5}{4\sqrt{2}\pi} \frac{\d^2}{\d p_z\d p_i} \int
d^3x\  e^{-i\v p\v x}\frac{e^{-3/2\a r}}{r},
\end{equation}
and consequently
\begin{equation}
  \chi_i(\v p)={\a^5}{\sqrt{2}}\left[\frac{4\ p_i\ p_z}{(p^2+9/4\ \a^2 )^3}-
  \frac{\de_{iz}}{(p^2+9/4\ \a^2 )^2}\right].
\end{equation}
The first term in the square bracket is longitudinal (proportional to
$p_i$) and will be canceled by the projection on the transversal
directions in the two-point function.

\subsection{Spontaneous emission in the presence of vacuum}

We now compute the kernel of the equation \eqref{main_eq} for the $2P\ra
1S$ transition ($\w=\a^2(\tfrac{1}{2}-\tfrac{1}{8})$ in dimensionless
units) in the presence of the vacuum state. The smeared two-point
function, $S(t-s)$, of equation \eqref{vac_2_point} will be employed.
After integrating out the angle variables, we get
\begin{equation}
S(t-s)=\frac{\a^2}{3\pi^2}\ \int_0^\infty \frac{p\;
dp}{[(p/\a)^2+9/4]^4}\ e^{-ip(t-s)}.
\end{equation}
From this point, the investigation of the spontaneous emission process,
in particular of the type of decay-curve, and of the life-time of the
$2P$ state, can be addressed with the help of mathematical methods for
Volterra equations \cite{volt}.

\section{Conclusions and outlook}
 We have presented a simple model of the spontaneous emission of light.
While the restriction to the one-excitation (photon) transitions
is the most obvious limitation of this model, it offers, on the
other hand, a generalization with respect to the allowed initial
states of the quantum electromagnetic field. It asserts that the
characteristics of the emission process are not intrinsic or
unique properties of atoms, but rather they depend on the initial
state of the quantum electromagnetic field\footnote{This point is
obvious for example from the explicit results of \cite{bon}.}.
Two future lines of research seem interesting: firstly one might
ask, whether the emission process in the distant (cosmological)
past was similar to the one we observe in laboratories nowadays.
The two-point function of the initial state of the quantum
electromagnetic field, at that epoch, was certainly different
from the (Minkowski-) vacuum two-point function. The second
direction, which we regard as promising, is to endeavor to use
simple quantum systems (e.g. some probe-atoms) as detectors of
subtle non-classical properties of quantum electromagnetic
fields. The temporal dependence of the excited-state's amplitude,
if measured with sufficient resolution, might reveal fingerprints
of the, for instance, negative energy-densities of the field.

\begin{acknowledgement}
The authors would like to thank prof.~K.~Fredenhagen for his
remarks on the structure of the model, as well as for drawing our
attention to the works of V.~Bach. One of the authors (P.~M.)
acknowledges the financial support of DFG and the hospitality of
DESY.
\end{acknowledgement}

\appendix
\section{Dimensionless units} \label{appa}
In this paper we use dimensionless units. They are defined as follows:
\begin{align*}
E&= mc^2\ E_D, & \v x&=\dfrac{\hbar}{mc}\ \v x_D,\\
t&=\dfrac{\hbar}{mc^2}\ t_D, & &p= mc\ p_D,\\
\psi&=\psi_D\ \left(\frac{\hbar}{mc}\right)^{-3/2},&
A&=A_D\sqrt{\dfrac{m^2c^3}{\hbar}},
\end{align*}
where the subscript $D$ denotes the dimensionless
quantities\footnote{As in the paper we use only the dimensionless
expressions, this subscript is dropped for brevity.}. The only
constant which can appear in the equations is the fine-structure
constant $\a$. If we take the interaction term as in the equation
\eqref{equ:interaction}, then the quantities corresponding to the
free electromagnetic field, in particular: the two-point
function, should not contain $\a$. As the unit of time is
particularly important in the context of this paper, we note that
$t_D=1$ corresponds to $\hbar/ m_e c^2 \approx 1.25\*10^{-21}$s.

\bibliographystyle{plain}


\end{document}